\documentclass[letter,12pt]{article}
\usepackage{epsfig}
\usepackage{tabularx}
\usepackage{amsmath}
\begin{document}
\renewcommand{\thefootnote}{\fnsymbol{footnote}}
\begin{flushright}
OSU-HEP-09-06
\end{flushright}
\begin{center}
{\Large Fermion Masses and Mixings in a Minimal \\$SO(10)\times A_4$
SUSY GUT} \\ \vspace{1cm}
{\large Abdelhamid Albaid\footnote{\texttt{Email:abdelhamid.albaid@okstate.edu}}} \\
\vspace{0.3cm} {\it Department of Physics, Oklahoma State
University, Stillwater, OK, 74078 USA} \vspace{0.1cm}(September 23,
2009)

\end{center}
\begin{abstract}
A SUSY SO(10) $\times A_4$ GUT model is constructed for fermion
masses and mixing by introducing a minimal set of low dimensional
Higgs representations needed to break the guage symmetry down to
$SU(3)_c \times U(1)_{em}$. The hierarchy of fermion masses can be
understood in the framework of $A_4$ symmetry. From the
$A_4$-invariant superpotential, the ``double lopsided'' mass
matrices for the charged leptons and the down quarks are obtained.
It is shown that this structure leads to bi-large neutrino mixings
simultaneously with small CKM mixing angles. An excellent fit to the
 masses and mixings of the quarks and leptons as well as to CP violation parameter
is obtained. Moreover, the model predicts the neutrino mixing angle
$\sin\theta_{13}\approx 0.15$.
\end{abstract}

\renewcommand{\thefootnote}{\arabic{footnote}}\setcounter{footnote}{0}

\section{Introduction}
\setlength{\baselineskip}{22pt} There are many unification models
that try to combine the strong and electroweak interactions into a
simple group. The simplest model is based on SU(5) gauge symmetry
\cite{a}. The minimal SU(5) model predicts the good mass relation
for the third generation (i.e. $m^0_b=m^0_{\tau}$) at GUT scale.
However, it has a bad prediction for mass relations for the first
and second generations (i.e. $m^0_s=m^0_{\mu}$, $m^0_d=m^0_{e}$) at
GUT scale. In addition, it does not naturally accommodate the right
handed neutrino. On the other hand, the SO(10) model accommodates
all chiral fermions of one standard model generation plus a right
handed neutrino within a 16-dimensional irreducible representation
(irrep.). Also, minimal SO(10) with only $10_H$ leads to the up
quark mass matrix being proportional to the down quark mass matrix,
so it is considered a good zeroth order approximation for CKM
mixings. Some models based on SO(10) symmetry without including any
family symmetry were proposed to explain most of the features of
quarks and leptons \cite{b}\cite{z}. However, one is not really
fully satisfied with only producing the fermion masses and mixing
angles without explaining why we have three generations and without
understanding the relation among generations, such as hierarchy and
mixing angles. For example, the flavor symmetry $A_4$
\cite{c}\cite{c1} can be employed to explain why the observed
neutrino mixing matrix is in very good agreement with the so called
tri-bi-maximal(TBM) mixing structure \cite{d}. Thus, it may be
important to consider the underlying family symmetry. One of the
best candidates for flavor symmetry is the non-Abelian discrete
symmetry $A_4$, for the following reasons. Firstly, it is the
smallest group that has a 3-dimensional irrep. Secondly,
SUSY-SO(10)$\times$ $A_4$ symmetry solves FCNC problem since the
scalar fermions, which belong to the 16-irrep of SO(10) and
transform as a triplet under $A_4$, have degenerate masses. Finally,
it was shown that the TBM mixing structure can be obtained by
imposing $A_4$ symmetry \cite{c1}.

Few models based on the SO(10) $\times$ $A_4$ group have been
studied \cite{e}\cite{f}. In these models, large Higgs
representations have been employed. For example, in ref.\cite{e},
the authors employed a ($126_H$,3) representation, where the first
(second) entry indicates the transformation under SO(10) ($A_4$), in
order to produce the fermion masses and mixing angles for both
normal and inverted neutrino mass spectra. Besides employing the
large Higgs representation $126_H$, the models in ref.\cite{f}
contain more than one adjoint $45_H$ representation. It has been
shown that only one adjoint Higgs field is required to break SO(10)
while preserving the gauge coupling unification \cite{g}. Also,
using quite large Higgs representation like $126_H$ leads to the
unified gauge coupling being non-perturbative before the Planck
scale, which might be hard to obtain from superstring theory
\cite{h}. Therefore, the purpose of this paper is to construct
SO(10) $\times$ $A_4$ model in which SO(10) is broken to the
standard model (SM) group in the minimal breaking scheme. This means
using only a spinor-antispinor ($16_H$,$\overline{16}_H$) to break
the rank of SO(10) from five to four, and the right handed neutrino
gets a heavy mass from the antispinor Higgs field
($\overline{16}_H$). Then one adjoint representation $45_H$ is used
to break the group all the way to the SM group. Recently, a
numerical analysis for quark and charged lepton masses and mixings
based on non-supersymmetric SO(10) without flavor symmetry was done
\cite{z}. The authors did not include the neutrino sector in the
numirical fitting. Their result for atmospheric angle was
$\sin\theta_{atm}=0.89$. However, as this paper shows, when the
neutrino sector is included, the result is not only a better fit for
atmospheric angle $\sin\theta_{atm}=0.776$, but also the known light
neutrino mass differences are accommodated.

The paper is organized as follows. In section 2, a general structure
of the fermion mass matrices for the second and third generations is
constructed. Then, based on that structure, the fermion mass
hierarchy and relations are explained. In section 3, it is shown
that introducing several 10-plets of matter fields to the model
leads to the doubly lopsided structure which produces large neutrino
mixing angles and small quark mixing angles simultaneously \cite{i}.
Then, some analytical expressions for quark masses and mixing angles
at GUT scale are derived in a certain approximation to the model
parameters. In section 4, exact numerical analysis is done to find
the outputs at GUT scale. To get predictions of fermion masses and
mixings at low scale, the quark masses and mixing at GUT scale will
be run to the low scale  by using renomalization equations. In
section 5, It is shown how to get a suitable right-handed neutrino
mass structure that gives the correct fitting for atmospheric angle
after combining the charged lepton contribution. Section 6 is the
conclusion.
\section{Fermion Mass Structure in $SO(10)\times A_4$ Symmetry}
In this section, the renormalizable Yukawa couplings of the SM
fermions with the extra spinor-antispinor matter fields are
considered as a concrete example of the model. The known matter
fields of the SM (quarks and leptons) plus the right handed neutrino
are contained in the three spinors ($16_i$,3), where $i$ stands for
flavor index. The ordinary fermions, $16_i$, do not couple with
$45_H$ in the minimal SO(10). As a result, some of the predictions
of minimal SO(10) such as $m_{\mu}=m_s$ and $m_c/m_t=m_s/m_b$ will
follow, which are badly broken in nature. Therefore, extra heavy
fermion fields must be introduced in order to allow the $45_H$ to
couple directly with the quarks and leptons of the standard model.
The transformation of the ordinary fermions and the extra matter
fields under $A_4$ and the additional symmetry $Z_2\times Z_4\times
Z_2$ are summarized in Table 1. Let us consider first the invariant
superpotential $W_1$ under the assigned symmetry that contains the
coupling of ordinary fermions with the spinor-antispinor matter
fields.
\begin{eqnarray}
W_1 &=& b_1 16_i \overline{16}_1 1_{Hi}+ b_2 16_i \overline{16}_2 1'_{Hi} +
\Omega 16_1 \overline{16}_3 45_H+ a 16_3 16_2 10_H  \nonumber \\
 &&+ M_1 16_1 \overline{16}_1 + M_2 16_2 \overline{16}_2 + M_3 16_3
 \overline{16}_3.
\end{eqnarray}
Table 2 summarizes the transformation of the Higgs fields needed to
achieve a minimum breaking scheme and the Higgs singlets needed to
break $A_4$ symmetry. Although in this model, the structure in Eq.1
does not include the Yukawa term $16_i 16_i 10_H $ which is
forbidden by the discrete symmetry $ Z_2 \times Z_4 \times Z_2 $,
the ordinary standard model fermions get their masses through their
coupling with heavy extra fields. This is similar to how the light
neutrinos get their masses through the coupling with the heavy right
handed neutrinos in the known see-saw mechanism.

The coupling terms in the superpotential $W_1$ can be represented
diagrammatically as shown in Fig.1. After integrating out the heavy
states, the approximate effective operators can be read from the
diagram.
\begin{eqnarray}
W_{ij} \approx \sum_{ij} \frac{16_i 16_j \langle45_H\rangle
\langle10_H\rangle \langle1_{Hi}\rangle \langle 1'_{Hj}\rangle }{M_1
M_2 M_3}.
\end{eqnarray}
The VEVs of Higgs fields can be written down in general form as:
\begin{eqnarray}
\langle45_H\rangle = \Omega  Q , \\
\langle1_{Hi}\rangle =  \left( \begin{array}{c} \epsilon_1 \\  \epsilon_2 \\
\epsilon_3
\end{array} \right), \\
\langle1'_{Hi}\rangle =  \left( \begin{array}{c} s_1 \\  s_2 \\
 s_3
\end{array} \right ),\\
\langle5(10)\rangle= v_u    ,   \langle\overline{5}(10)\rangle= v_d.
\end{eqnarray}
Here the notation $\langle p(q)\rangle$ refers to a $p$ of SU(5)
contained in a $q$ of SO(10). The $Q$ from Eq.3 is a linear
combination of SO(10) generators. One can redefine, without loss of
generality, the light fermion states as:
\begin{table}
\center
\begin{tabular}{|c|c|c|c|c|c|c|} \hline
 SO(10)                     &$16_i$     &$16_1$,$\overline{16}_1$   &$16_2$,$\overline{16}_2$   &$16_3$,$\overline{16}_3$   &$1^c_i$    \\\hline
 $A_4$                      & 1         & 1                         & 1                         & 1                         &3          \\\hline
 $Z_2\times Z_4\times Z_2$  &+,+,+      &+,-,+                      & -,+,+                     &+,+,-                      &+,+,+      \\\hline
 SO(10)                     &$10_i$     & $10'_i$                   &$10''_i$                   &$10'''_i$                  &$1_i$      \\\hline
 $A_4$                      &3          &3                          &3                          &3                          &3          \\\hline
 $Z_2\times Z_4\times Z_2$  &+,$i$,+    &+,$-i$,+                   &+,$i$,-                    &+,$-i$,-                   &+,$-i$,+   \\\hline
\end{tabular}
\caption{The transformation of the matter fields under
SO(10)$\times$$A_4$ and $ Z_2 \times Z_4 \times Z_2 $.}
\end{table}
\begin{eqnarray}
16_1 \epsilon_1 + 16_2 \epsilon_2 + 16_3 \epsilon_3 &=& \epsilon
16'_3 \nonumber\\ 16_1 s_1 + 16_2 s_2 + 16_3 s_3
&=&S(16'_2s_{\theta}+16'_3c_{\theta}),
\end{eqnarray}
where $\epsilon=\sqrt{\epsilon^2_1+\epsilon^2_2+\epsilon^2_3}$ and
$S=\sqrt{s^2_1+s^2_2+s^2_3}$. In terms of the redefined light
fermion states, after dropping the prime notation and plugging in
the VEVs, one gets
\begin{eqnarray}
W_0 \approx\frac{\Omega \epsilon S \langle10_H\rangle }{M_1 M_2
M_3}(16_3 16_2 Q_{(16_3)} s_{\theta} +16_3 16_3 Q_{(16_3)}
c_{\theta}).
\end{eqnarray}
In general, the above effective operator can be written in terms of
quark and lepton fields as
\begin{eqnarray}
W_F \approx\frac{\Omega \epsilon S \langle10_H\rangle }{M_1 M_2
M_3}(F_3 F^c_2 Q_F s_{\theta}+F^c_3 F_2 Q_{F^c} s_{\theta} +F_3
F^c_3( Q_F + Q_{F^c}) c_{\theta}).
\end{eqnarray}
Here $F$ is a general notation for up quarks ($U$), neutrinos ($N$),
charged leptons ($L$) and down quarks ($D$). The quantity
$Q_F$$(Q_{F^c})$ refers to the assigned charge of the left handed
fermion (charge conjugate of the left handed fermions) after
breaking SO(10) group down to the the SM group. The unbroken charge
Q can be written  as a linear combination of two generators that
commute with $SU(3)_c\times SU(2)_L\times U(1)_Y$ as follows:
\begin{eqnarray}
Q=2I_{3R}+ \frac{6}{5}\delta(\frac{Y}{2}),
\end{eqnarray}
where $I_{3R}$ is the third generator of $SU(2)_R$ and Y is the
hypercharge of the Abelian U(1) group. The charge Q for different
quarks and leptons is given below.
\begin{eqnarray}
Q_u=Q_d=\frac{1}{5}\delta, \hspace{0.5cm}  Q_{u^c}=-1-\frac{4}{5}\delta, \hspace{0.5cm} Q_{d^c}=1+\frac{2}{5}\delta, \nonumber \\
Q_{l}=Q_{\mu}=-\frac{3}{5}\delta, \hspace{0.5cm}
Q_{l^c}=1+\frac{6}{5}\delta, \hspace{0.5cm} Q_{\nu^c}=-1.
\end{eqnarray}
Eq.10 can be expressed in the following matrix form:
\begin{eqnarray}
W_F \approx \left(
\begin{array}{ccc}
F^c_1 & F^c_2 & F^c_3
\end{array} \right) ( \frac{\Omega \epsilon S \langle10_H\rangle }{M_1 M_2 M_3})  \left(
\begin{array}{ccc}
0 & 0 & 0 \\
0 & 0 & Q_{F} s_{\theta} \\
0 & Q_{F^c} s_{\theta} & ( Q_F + Q_{F^c}) c_{\theta}
\end{array} \right) \left(\begin{array}{c} F_1 \\ F_2 \\ F_3
\end{array}\right).
\end{eqnarray}
It is assumed that the masses of the extra heavy fermions are a
little above the GUT scale,  so the factors $N_F=(1+T^2Q^2_F)^{1/2}$
that come from doing the algebra exactly are approximately equal to
one, where $T=\Omega \epsilon S /M_1 M_2 M_3$. The first feature of
the general mass matrix of the light fermions in Eq.13 is an
explanation for the mass hierarchy between the second and third
generation in the limit $s_{\theta}\rightarrow 0$. It is remarkable
that a relation among generations is related to the vacuum alignment
of $A_4$ Higgs.

Another feature of the above light fermion mass matrix
$m^0_b=m^0_{\tau}$ is obtained through $M_{D33}=M_{L33}$, which
follows from the relation $Q_{d^c}+Q_d=Q_{l^c}+Q_l$. This relation
occurs because both down quarks and charged leptons get their masses
from the same Higgs.

A further consequence of the light fermion mass structure is that
$m^0_s\neq m^0_{\mu}$. This inequality relation follows from
$m^0_{\mu}/m^0_s=L_{32}L_{23}/D_{32}D_{23}=Q_{l^c}Q_{l}/Q_{d^c}Q_{d}$
which is not necessarily equal to one. This leads to the following
question: What VEV direction should be given to $45_H$ in order to
obtain the Georgi-Jarlskog relation $m^0_{\mu}=3m^0_s$ ? There are
two choices, either $\delta\rightarrow 0$ or $\delta\rightarrow
-1.25$. The former choice gives the unwanted relation
$(m^0_c/m^0_t)/(m^0_s/m^0_{\tau})\rightarrow 1$, while the later
leads to $(m^0_c/m^0_t)/(m^0_s/m^0_{\tau})\rightarrow 0$. Thus, a
good fit for $\delta$ should be around -1.25.
\begin{table}
\center
\begin{tabular}{|c|c|c|c|c|c|c|c|c|c} \hline
 SO(10)&$10_H$ & $45_H$ & $16_H$& $\overline{16}_H$&$1_{Hi}$ & $1'_{Hi}$&$1''_{Hi}$&$1'''_{Hi}$\\ \hline
 $A_4$ & 1 & 1 & 1 & 1&3&3&3&3\\   \hline
 $Z_2\times Z_4\times Z_2$ &-,+,-&+,-,-& +,$-i$,+ & +,$-i$,+&+,-,+&-,+,+&+,+,-&+,$i$,+\\   \hline
\end{tabular}
\caption{The transformation of the Higgs fields under
SO(10)$\times$$A_4$ and $ Z_2 \times Z_4 \times Z_2 $. }
\end{table}
\section{Extension to the First Generation and Doubly Lopsided Structure }
In this section, the vector 10-plets fermions are added to the model
to generate masses and mixings of the first generation. Adding these
vector multiplets does not contribute to the up quark mass matrix
since 10-plets do not contain a charge of ($\pm2/3$). Therefore, up
quark matrix is still rank 2 and this is consistent with $
\frac{m^0_u}{m^0_t}\approx10^{-5}$ being much smaller than $
\frac{m^0_d}{m^0_b}\approx10^{-3}$ and $
\frac{m^0_e}{m^0_{tue}}\approx 0.3\times 10^{-5}$. First, it will be
shown how the model leads to the doubly lopsided structure by
employing these vector multiplets, then some analytical expressions
for masses and mixing angles of fermions at GUT scale will be
derived. Let us first consider the invariant couplings under the
assigned symmetry which can be read from the Feynman diagram in
Fig.2. The allowed couplings in the superpotential $W_{2}$ are
\begin{eqnarray}
W_{2}=
16_i10_i16_H+M_{10}10_i10'_i+h'_{ijk}10'_i10'_j1_{Hk}+h_{ijk}10_i10_j1_{Hk}.
\end{eqnarray}
The important point is that Fig.2 gives a flavor-symmetric
contribution to the down quark and charged lepton mass matrices. In
order to understand this, recall that the general product of three
triplets: ($a_1$, $a_2$, $a_3$), ($b_1$, $b_2$, $b_3$), and ($c_1$,
$c_2$, $c_3$) that transform as a singlet under $A_4$ is given by
\begin{eqnarray}
 h_1(a_2b_3c_1+a_3b_1c_2+a_1b_2c_3)+h_2(a_3b_2c_1+a_1b_3c_2+a_2b_1c_3).
\end{eqnarray}
The third term of Eq.13 gives a symmetric contribution since there
are two identical 10-plets. The last term in Eq.13 has been ignored
by assuming
 the Yukawa couplings $h_{ijk}$ are very small. The contribution of Fig.2 to the mass matrices of the down quarks and charged leptons after integrating out the extra vector multiplets is then
\begin{eqnarray}
 M^s_L=M^s_D\propto\left(
\begin{array}{ccc}
0 & c_{12} & c_{13} \\
c_{12} & 0 & c_{23} \\
c_{13} & c_{23} & 0
\end{array} \right),
\end{eqnarray}
where $c_{12} $, $c_{13} $, and $c_{23}$ are proportional to
$\epsilon_1$, $\epsilon_2$, $\epsilon_3$ respectively. To obtain the
desired fermion mass structure (the doubly lopsided structure, which
is going to be explained later in this section), other couplings are
needed to be included by employing the four vector 10-plets plus
adding another Higgs singlet $1''_{iH}$ to the model (their
transformations under the assigned symmetry are shown in Tables 1
and 2). The purpose of these couplings is to give a
flavor-antisymmetric contribution to the down quarks and charged
leptons mass matrices. Since the adjoint of SO(10) ($45_H$) is an
antisymmetric tensor which changes the sign under the interchange
$10'_i\leftrightarrow10'''_i$. One can consider employing the Yukawa
coupling $10'''_i10'_i45_H$. Also, from the fact that when we write
the SO(10)-vectors in the SU(5) basis such as
$10_i=5_i+\overline5_i$, the charged lepton and down quark content
of $5_i$ or $\overline5_i$ have different chirality, therefore the
structures of matrices $M_L$ and $M_D$ have opposite signs (look at
the mass structures in Eqs.18-19). It is important to emphasize that
the minimum Higgs breaking scheme assumption does not allow us to
add another adjoint to the model. Therefore, the same adjoint $45_H$
Higgs representation that breaks SO(10) group to the SM group is
going to be used. Additional couplings to the previous
superpotential can be read from Fig.3.
\begin{eqnarray}
W_{3}=10'_i10''_j1''_{Hk}+m10''_i10'''_i+10'''_i10'_i45_H,
\end{eqnarray}
where $\langle45_H\rangle$ has been defined previously. The
following VEV to the Higgs singlet $1''_H$ is given below:
\begin{eqnarray}
\langle1''_H\rangle=\left(\begin{array}{c}
\delta_1 \\ \delta_2 \\ \delta_3
\end{array} \right).
\end{eqnarray}
 After integrating out the heavy states, the
following contribution to the $M_L$ and $M_D$ is obtained:
\begin{eqnarray}
 M^A_L&\propto&\left(
\begin{array}{ccc}
0 & -\delta_3Q_l & \delta_2Q_l \\
\delta_3Q_l & 0 & -\delta_1Q_l \\-\delta_2Q_l & \delta_1Q_l & 0
\end{array} \right),\\
 M^A_D& \propto &\left(
\begin{array}{ccc}
0 & \delta_3Q_{d^c} & -\delta_2Q_{d^c} \\
-\delta_3Q_{d^c} & 0 & \delta_1Q_{d^c} \\
\delta_2Q_{d^c} & -\delta_1Q_{d^c} & 0
\end{array} \right),
\end{eqnarray}
where the overall constant has been absorbed in the redefinition of
$\delta_1$, $\delta_2$ and $\delta_3$. Eqs.18-19 show that the
off-diagonal elements of $M^A_D$ ($M^A_L$) are proportional to
$Q_{d^c}$ ($Q_l$). This is because $\overline5_i(10)$ contains in
its representation the charge conjugation of a color triplet of the
left handed down quarks $d^c_{Li}$ and the left handed charged
leptons $e_{Li}$. The full tree level mass matrices, which are
obtained by adding the three superpotentials $W_1+W_{2}+W_3$, have
the following forms:
\begin{eqnarray}
 M_L &=& m^0_d\left(
\begin{array}{ccc}
0 & c_{12}+3\delta_3(\frac{-1+\alpha}{5}) & -\delta_2\alpha+\zeta \\
c_{12}-3\delta_3(\frac{-1+\alpha}{5}) & 0 & \delta_1\alpha-3s(\frac{-1+\alpha}{5})+\beta \\
\zeta-\delta_2\frac{6-\alpha}{5} & s(\frac{-1+6\alpha}{5})+\delta_1(\frac{6-\alpha}{5})+\beta & 1
\end{array} \right),\\
 M_D &=& m^0_d\left(
\begin{array}{ccc}
0 & c_{12}+\delta_3(\frac{3+2\alpha}{5}) & -2\delta_2(\frac{3+2\alpha}{5})+\zeta \\
c_{12}-\delta_3(\frac{3+2\alpha}{5}) & 0 & 2\delta_1(\frac{3+2\alpha}{5})+s(\frac{-1+\alpha}{5})+\beta \\
\zeta & s(\frac{3+2\alpha}{5})+\beta & 1
\end{array} \right),\\
 M_U&=&m^0_u\left(
\begin{array}{ccc}
0 & 0 & 0 \\
0 & 0 & (\frac{1-\alpha}{5})s \\0 & (\frac{1+4\alpha}{5})s& 1
\end{array} \right),\\
 M_N &=&m^0_u\left(
\begin{array}{ccc}
0 & 0 & 0 \\
0 & 0 & (\frac{-3+3\alpha}{5})s \\0 & s & 1
\end{array} \right),
\end{eqnarray}
where the convention being used here is the left handed fermions
multiplied from the right. The parameters of the model have been
defined as follows:
\begin{eqnarray}
\zeta &=&c_{13}+\delta_2Q_{d^c},\nonumber\\\beta &=&
c_{23}+\delta_1Q_{d^c},\nonumber\\\delta
&=&-1+\alpha,\\s&=&\frac{s_{\theta}}{(\frac{3}{5}\delta+1)c_{\theta}}\nonumber .
\end{eqnarray}
The above fermion mass structures have eight parameters. If $\alpha$
goes to zero, the fermion mass matrices in Eqs.20-21 go to the SU(5)
limit ($m^0_b=m^0_{\tau}$, $m^0_s=m^0_{\mu}$, $m^0_d=m^0_{e}$). To
avoid the bad prediction of SU(5) for lighter generations, the good
numerical fitting for $\alpha$ should deviate from zero. On the
other hand, to keep the good SU(5) prediction for third generation,
the parameter $\alpha$ should satisfy $ \alpha<<1$. If $\delta_1$
and $\delta_2$ are of order one and the other model parameters are
very small ($\beta,\zeta,\alpha,\delta_3,c_{12},s<<
\delta_1,\delta_2$), the model leads to the doubly lopsided
structure. To see this clearly, let us go to the limit where the
small parameters are zero (except $s$). So the $M_D$ and $M_L$ go to
the following form:
\begin{eqnarray}
 M_L=M^{T}_D=m^0_d\left(
\begin{array}{ccc}
0 & 0 & 0\\
0 & 0 & (\frac{3s}{5}) \\
-\delta_2\frac{6}{5} & (\frac{-s}{5})+\delta_1(\frac{6}{5}) & 1
\end{array} \right).
\end{eqnarray}
In diagonalizing $M_L$ of the Eq.25, the large off diagonal elements
$\delta_1$ and $\delta_2$ that appear asymmetrically in $M_D$ and
$M_L$ must be eliminated from the right by a large left handed
rotation angle $\theta_{sol}$ in the 1-2 plane, where
$\tan(\theta_{sol})=-\frac{\delta_2}{\delta_1}$. The next step of
diagonalization is to remove the large element
$\sigma\approx(\delta^2_1+\delta^2_2)^\frac{1}{2}$  that has been
produced after doing the first diagonalization where (3,2) element
of the matrix in Eq.25 is replaced by $\sigma$. This can be done by
a rotation acting from the right by a large left handed angle
$\theta_{23}$ in the 2-3 plane, where
$\tan(\theta_{23})\approx-\sigma$. On the other hand, there are no
corresponding large left handed rotation angles in diagonalizing
$M_D$ since $M_L=M^{T}_D$. However, the large off diagonal elements
in $M_D$ can be eliminated by large right handed rotation angles
acting from the left on the $M_D$ in Eq.25, while the left handed
rotation angles are small. This explains how the doubly lopsided
structure leads to the small CKM mixing angles and the large
neutrino mixing angles simultaneously. If the parameters $c_{12}$,
$\delta_3$, and $\zeta$ are zero, analytical expressions can be
written down for the ratios of quark and lepton masses of the second
and third generations, $V_{cb}$, and neutrino mixing angles (
$\tan\theta_{12}$ and $\tan\theta_{23}$) in terms of $\delta_1$,
$\delta_2$, $s$ ,$\alpha$, and $\beta$:
\begin{eqnarray}
\frac{m^0_c}{m^0_t}&=&\frac{s^2(1-\alpha)(1+4\alpha)}{25}\nonumber,\\
\frac{m^0_s}{m^0_b}&=&\frac{-2(3+2\alpha)(\beta+s(\frac{3+2\alpha}{5}))\sqrt{\delta^2_1+\delta^2_2}}{5(1+\frac{4}{25}(3+2\alpha)^2(\delta^2_1+\delta^2_2))}\nonumber,\\
\frac{m^0_{\mu}}{m^0_{\tau}}&=&\frac{
\sqrt{(\frac{-3s}{5}(-1+\alpha)+\delta_1\alpha+\beta)^2+\delta^2_2\alpha^2}\sqrt{(\delta^2_1+\delta^2_2)}(6-\alpha)}{5(1+\frac{(6-\alpha)^2}{25}(\delta^2_1+\delta^2_2))},\\
V^D_{cb}&=&\frac{\beta+\frac{s(3+2\alpha)}{5}}{(1+\frac{4}{25}(3+2\alpha)^2(\delta^2_1+\delta^2_2))}\nonumber,\\
V^U_{cb}&=&\frac{-s(1+4\alpha)}{5}\nonumber,\\
\tan\theta_{12}&=&\frac{\delta_2(\frac{6-\alpha}{5})}{\delta_1(\frac{6-\alpha}{5})+s(\frac{-1+6\alpha}{5})+\beta},\nonumber\\
\tan\theta_{23}&=&-(\frac{6-\alpha}{5})\sqrt{\delta^2_2+\delta^2_1}\nonumber.
\end{eqnarray}
These expressions are derived by using the approximation $\alpha,s,
\beta<<\delta_1,\delta_2$ and are useful for fitting the data. The
best fit for the data is obtained by setting $\tan\theta_{23}=-2$
and $\tan\theta_{12}=0.68$, which correspond to $\theta_{23}=-63^o$
and $\theta_{12}=34^o$. The central value of the atmospheric angle
is around $45^o$. In order to bring $63^o$ down close to the central
value, the neutrino sector is required to be included as shown in
section 5. Also, it will be shown that the contribution of the
neutrino sector to the solar angle is small.
\section{The Numerical Results}
The model can be shown to be concrete by giving numerical values to
the parameters of the model, and producing the six mass ratios of
quarks and leptons, CKM mixing angles ($V_{us}$, $V_{ub}$, and
$V_{cb}$), CP violation parameter $
\eta=-Im(V_{ub}V_{cs}/V_{us}V_{cb})$, and neutrino mixing angles ($
\sin\theta_{12}$, and $ \sin\theta_{13}$). The fermion mass texture
in Eqs.20-23 has eight parameters ($\delta_1$, $\delta_2$,
$\delta_3$, $\alpha$, $\beta$, $s$, $\zeta$, and $ c_{12}$). These
parameters are going to be used to fit 12 quantities. If
$\delta_1=-1.302$, $\delta_2=1.0142$, $\delta_3=0.015\times
e^{4.95i}$, $\alpha=-0.05801$, $ s=0.29$, $\zeta=0.0105$, $
c_{12}=-0.00153e^{1.1126i}$, and $\beta=-0.12303$, the following
excellent fit at GUT scale is obtained :
$\frac{m^0_c}{m^0_t}=0.002717$, $\frac{m^0_b}{m^0_{\tau}}=0.958$,
$\frac{m^0_e}{m^0_{\mu}}=0.00473$,
$\frac{m^0_{\mu}}{m^0_{\tau}}=0.0585$, $\frac{m^0_d}{m^0_e}=3.63$,
$\frac{m^0_s}{m^0_{\mu}}=0.302$, $\eta=0.357$, $V_{us}=0.2264$,
$V_{ub}=0.0037$, $V_{cb}=0.0362$, $\sin\theta_{12}=0.569$, and
$\sin\theta_{13}=0.0653$. The above numerical fittings lead to
$\sin\theta^L_{23}=0.904$ which is not close to the central value
$\sin\theta^{atm}_{23}=0.707$. One can see from the superscript L
that the mixing angle $\theta^L_{23}$ comes only from the charged
lepton contribution. To obtain close to the expected atmospheric
angle and the correct neutrino mass differences, it is important to
include the neutrino sector contribution to the atmospheric angle by
finding out a suitable right handed neutrino structure which
respects the assigned symmetry of the model.
\begin{table}[t]
\center
\begin{tabular}{|c|c|c|c|c|c|c} \hline
 & models predictions & expt. & Pull\\ \hline
 $m_e(m_e)$ & 0.511$\times 10^{-3} $& 0.511$\times 10^{-3} $ & -  \\   \hline
 $m_{\mu}(m_\mu)$ & 105.6$\times 10^{-3} $ & 105.6$\times 10^{-3} $ &  - \\   \hline
 $m_{\tau}(m_\tau)$ & 1.776 & 1.776 & -  \\ \hline
 $\overline m_{ud}$ & $4.32\times 10^{-3}$ & ($3.85 \pm 0.52$)$\times 10^{-3}$ & 0.9 \\   \hline
 $m_c(m_c)$ & 1.4 & $1.27^{+0.07}_{-0.11}$ & 1.85 \\   \hline
 $m_t(m_t)$ & 172.5 & 171.3$\pm$2.3 & 0.52  \\   \hline
 $\frac {m_s}{\overline m_{ud}}$ & 25.36 & $27.3\pm1.5$ & 1.29 \\   \hline
 $m_s(2 Gev)$ & 109.6$\times 10^{-3}$ & $105^{+25}_{-35}\times 10^{-3} $ &  0.184 \\   \hline
 $m_b(m_b)$ & 4.31 & $4.2^{+0.17}_{-0.07}$ &  0.58  \\   \hline
 $V_{us}$ & 0.2264 & 0.2255$\pm$0.0019 &  0.473 \\   \hline
 $V_{cb}$ & 39.2$\times 10^{-3} $ & (41.2$\pm$1.1)$\times 10^{-3} $ & 1.82 \\   \hline
 $V_{ub}$ & 4.00$\times 10^{-3} $ & (3.93$\pm$0.36)$\times 10^{-3} $& 0.194  \\   \hline
 $\eta$ & 0.3569 & $0.349^{+0.015}_{-0.017}$ &0.526 \\   \hline
 $sin\theta^{sol}_{12}$ & 0.551* & 0.566$\pm$0.018 & 0.83 \\   \hline
 $sin\theta^l_{23}$ & 0.776* & 0.707$\pm$0.108 & 0.63 \\   \hline
 $sin\theta_{13}$ & 0.154* & $<0.22$ & - \\   \hline
\end{tabular}
\caption{This Table shows the comparison of the model predictions at
low scale and the experimental data.}
\end{table}

In order to compare with experiment,  the predicted fermion masses
and mixing angles at the low energy scale are needed to be found.
The above numerical values of the fermion masses and mixing angles
which are obtained at GUT scale have been evolved to the low scale
in two steps. First, the running from GUT scale to $M_{SUSY}=1
\textrm{ TeV}$ is done by using the 2-loop MSSM beta function. The
running factors denoted by $\eta_i$ depend on the value of
$\tan\beta$. The known fermion masses and mixing data are best
fitted with $\tan[\beta]=10$. The running factors for
$\tan[\beta]=10$ are ($\eta_{s/b}$, $\eta_{\mu/ \tau}$, $\eta_{b/
\tau}$, $\eta_{c/t}$, $\eta_{cb}$= $\eta_{ub}$)=(0.8736, 0.9968,
0.5207, 0.73986, 0.910335), where $\eta_{i/j}=(m^0_i/m^0_j)/(m_i(1
Tev)/m_j(1 Tev))$ and $\eta_{cb,ub}=V^0_{cb,ub}/V_{cb,ub}(1 Tev)$.
The second step is to evolve the fermion masses and mixing angles
from $M_{SUSY}=1 \textrm{ TeV}$ to the low scale. The
renormalization factors $\eta_i$ that run fermion masses from their
respective masses up to the supersymmetric scale $M_{SUSY}=1
\textrm{ TeV}$ are computed using 3-loop QCD and 1-loop QED, or
electroweak renormalization group equation (RGE) with inputs
$\alpha_s(M_Z)=0.118$, $\alpha(M_Z)=1/127.9$ and
$\sin\theta_w(M_Z)=0.2315$. The relevant renormalization equations
can be found in \cite{j}\cite{j1}. The results are ($\eta_c$,
$\eta_b$, $\eta_{e}$, $\eta_{\mu}$, $\eta_{\tau}$, $\eta_t$,
$\eta_{ub}$=$\eta_{cb}$)=(0.4456, 0.5309, 0.8188, 0.83606, 0.8454,
0.98833, 1.0151).

By using the above renormalization factors, $m_{\tau}=1776 \textrm{
MeV}$, and $m_t=172.5 \textrm{ GeV}$, the following predictions at
low scale can be obtained: $m_c(m_c)= 1.4 \textrm{ GeV}$, $m_b(m_b)=
5.2 \textrm{ GeV}$, $m_e(m_e)=0.511 \textrm{ MeV}$,
$m_{\mu}(m_{\mu})= 105.6 \textrm{ MeV}$, $m_d(2\textrm{ GeV})=7.5
\textrm{ MeV}$, $m_s(2 \textrm{ GeV})= 132 \textrm{ MeV}$,
$\eta=0.357$, $V_{us}=0.2264$, $V_{ub}=0.004$, $V_{cb}=0.0392$,
$\sin\theta_{12}=0.569$, and $\sin\theta_{13}=0.0653$.

Note that the numerical value of $m_b$ is not in agreement with the
experimental value $m_b=4.20^{+0.17}_{-0.07}\textrm{ GeV}$ \cite{l}.
In order to fix this, the finite gluino and chargino loop
corrections \cite{k} are required to be included to the down type
quark masses ( $m_d$, $m_s$, $m_b$ ), which are denoted respectively
by (1+$\Delta_d$), (1+$\Delta_s$), (1+$\Delta_b$). These corrections
are proportional to the supersymmetric particle spectrum:
 $\Delta_b\approx
\tan\beta(\frac{2\alpha_3}{3\pi}\frac{\mu M_{\tilde g}}{m^2_{\tilde
b_L}-m^2_{\tilde b_R}}[f(m^2_{\tilde b_L}/M^2_{\tilde
g})-f(m^2_{\tilde b_R}/M^2_{\tilde
g})]+\frac{\lambda^2_t}{16\pi^2}\frac{\mu A_t}{m^2_{\tilde
t_L}-m^2_{\tilde t_R}}[f(m^2_{\tilde t_L}/\mu ^2)-f(m^2_{\tilde
t_R}/\mu^2)]) $, where $f(x)=\ln(x)/(1-x)$ and the first (second)
term refers to gluino (chargino) correction. Similar expressions
exist for $\Delta_s$ and $\Delta_d$, but without the chargino
contribution and $\tilde b\rightarrow \tilde s,\tilde d$. If the
chargino loop corrections are negligible and $m_{\tilde d}$,
$m_{\tilde s}$, and $m_{\tilde b}$ are degenerate, the equality
relation $\Delta_{d}=\Delta_{s}=\Delta_{b}$ is approximately
satisfied. In order to get better fitting for down type quark
masses, let us take $\Delta_{d}=\Delta_{s}=\Delta_{b}=-0.17$ which
gives $m_d(2\textrm{ GeV})= 6.24 \textrm{ MeV}$, $m^0_s(2 \textrm{
GeV})= 109.65 \textrm{ MeV}$, and $m_b(m_b)= 4.31 \textrm{ GeV}$.
The comparison of the model predictions and experimental data at low
scale is summarized in Table 3, where the quark and charged lepton
masses, the CKM mixing angles ($V_{ub}$, $V_{us}$, $V_{cb}$), the
neutrino mixing angles ($\sin\theta_{sol}$, $\sin\theta_{atm}$,
$\sin\theta_{13}$), and the CP violation parameter ($\eta$) are
taken from \cite{l}. The masses are all in GeV. Although the model
here predicts $m_u(GUT)=0$, the quantity $\overline
m_{ud}=(m_u+m_d)/2$ is considered in Table 3, where it is assumed
that the tiny up quark mass at GUT scale may be generated either by
including the coupling $16_i16_i10_H$ into the model or by
considering higher dimensional operators. Suppose $m_u(2\textrm{
GeV})=2.4 \textrm{ MeV}$,  the model predictions of the quantities
$\overline m_{ud}$ and $\frac {m_s}{ \overline m_{ud}}$, which are
well-known from lattice calculation \cite{m}, are given in Table 3.
The asterisks in Table 3 indicate that the model predictions of
neutrino mixing angles are obtained after including the neutrino
sector in section 5.
\section{Right Handed Neutrino Mass Structure}
Up to now, the model gives excellent agreement with the known values
for the CKM mixings, the quark masses, the charged lepton masses, CP
violation parameter, and the neutrino mixing angles
($\sin\theta_{12}$ and  $\sin\theta_{13}$). However, the whole
picture is still not complete and the following question arises.
What is the appropriate light neutrino mass matrix
($M_{\nu}=-M^T_NM^{-1}_RM_N$) that gives not only the correct
contribution to the atmospheric angle, but also the correct neutrino
mass differences: $\Delta m^2_{12}=(7.59 \pm 0.2)\times 10^{-5}
\textrm{ $eV^2$}$, $\Delta m^2_{32}=(2.42 \pm 0.13)\times 10^{-5}
\textrm{ $eV^2$}$ \cite{i}? In the other words, we are looking for a
suitable structure of right handed neutrino mass matrix $M_R$ since
$M_N$ is fixed. Recall that MNS mixing matrix is given by
\begin{eqnarray}
U_{MNS}=U^{\dag}_LU_{\nu},
\end{eqnarray}
where $U_L$ and $U_{\nu}$ are the unitary matrices needed to
diagonalize the hermitian lepton matrix $M^{\dag}_LM_L$ and the
light neutrino matrix $M_{\nu}$ respectively.
\begin{eqnarray}
M^{diag\dag}_LM^{diag}_L=U^{\dag}_LM^{\dag}_LM_LU_L, \hspace{0.5cm}
M^{diag}_{\nu}=U^{T}_{\nu}M_{\nu}U_{\nu},
\end{eqnarray}
where  $M_{\nu}$ is assumed to be real and symmetric. The Dirac
neutrino mass matrix $M_N$ in Eq.23 has vanishing first row and
column and the same is true for $M_{\nu}$. So the matrix required to
diagonalize $M_{\nu}$ is simply a rotation in the 2-3 plane by an
angle $\theta_{\nu}$ while $U^{\dag}_L$ is determined numerically
from the charged lepton mass matrix. Thus, the mixing matrix of
neutrinos is given by
\begin{small}
\begin{eqnarray}
 U_{MNS}=\left(
\begin{array}{ccc}
-0.1437-0.8068$i$ & 0.1353+0.5530$i$ & 0.0653\\
0.2497+0.0582$i$ & 0.3394-0.0400$i$ & 0.9041 \\
-0.5125+0.0002$i$ & -0.7477+0.00006$i$ & 0.4222
\end{array} \right)\left(
\begin{array}{ccc}
1& 0& 0\\
0 & \cos\theta_{\nu} & \sin\theta_{\nu} \\
0& -\sin\theta_{\nu} &\cos\theta_{\nu}
\end{array} \right).
\end{eqnarray}
\end{small}
One can conclude the correct contribution of the neutrino sector to
the atmospheric angle is around $\theta_{\nu}$=$-20^o$. For example,
if we take $\theta_{\nu}$=$-20^o$  the neutrino mixing angles
($\sin\theta_{atm}$, $\sin\theta_{sol}$, $\sin\theta_{12}$) become
 (0.707, 0.53, 0.21). In order to find the suitable
right handed neutrino mass structure, one can easily prove the
inverse of the see-saw relation.
\begin{eqnarray}
 M_R=-M_NU_{\nu}(M^{diag}_{\nu})^{-1}U^T_{\nu}M^T_{\nu}.
\end{eqnarray}
A similar technique was used in ref \cite{o}. Note that one of the
eigenvalues of $M_{\nu}$ is zero (i.e. $M^{diag}_{\nu}$ is
singular), so the inverse of $M^{diag}_{\nu}$ does not exist. To
overcome this problem, one can generally define
$M^{diag}_{\nu}$=diag( $m_1$, $m_2$, $m_3$ ) and $m_1$ will not
appear in $M_R$. By using the numerical result of $M_N$,
$\theta_{\nu}$=$-20^o$, and $m_2$/$m_3$=0.178, the right handed mass
structure can be presented numerically.
\begin{eqnarray}
\left(
\begin{array}{ccc}
0 & 0& 0\\
0& 0.0186 & -0.13 \\
0 & -0.13 & 1
\end{array} \right).
\end{eqnarray}
From the above numerical mass matrix, one concludes
$(M_R)_{23}\times(M_R)_{23}\approx(M_R)_{22}$, so to a good
approximation, the above numerical structure can be represented
analytically as follows:
\begin{eqnarray}
 \left(
\begin{array}{ccc}
0 & 0& 0\\
0& r^2 & ar \\
0 & ar & 1
\end{array} \right).
\end{eqnarray}
The constant $a$ should not be equal to one because then $M_R$ would
be singular, but the constant $a$ should be around 1. Now our
mission is to find the Yukawa couplings that respect the symmetry of
the model and lead to an analytical structure similar to Eq.32. This
can be accomplished by considering the following Yukawa couplings
represented by Feynmann diagram in Fig.4.
\begin{eqnarray}
W_4= 16_i\overline{16}_H1_i +h_{ijk}1_i1^c_j1'''_{Hk}+ m_11^c_i1^c_i.
\end{eqnarray}
Where two fermion singlets $1_i$ and $1^c_i$ which couple with the
singlet Higgs $1'''_{iH}$ have been introduced (their transformation
under SO(10)$\times$$A_4$ and the additional symmetry are shown in
Tables 1 and 2). The singlet transformation of the product of three
triplet under $A_4$ of the second term of Eq. 33 is given by
$h_1(N_1N^c_2\alpha_3+N_2N^c_3\alpha_1+N_3N^c_1\alpha_2)+h_2(N_1N^c_3\alpha_2+N_3N^c_2\alpha_1+N_2N^c_1\alpha_3)$,
where $\alpha_1$, $\alpha_2$, and $\alpha_3$ are the VEV's
components of $1'''_{iH}$. By assuming $h_1$=$h_2$, Fig.4 leads to
the desired right handed neutrino mass structure.
\begin{eqnarray}
M_R=\Lambda\left(
\begin{array}{ccc}
\frac{\alpha^2_1}{\alpha^2_3} & \alpha_1\alpha_2(\frac{-1}{\alpha^2_3}+\frac{2}{\alpha^2_1+\alpha^2_2+\alpha^2_3})& \frac{-\alpha_1(\alpha^2_1-\alpha^2_2+\alpha^2_3)}{\alpha_3(\alpha^2_1+\alpha^2_2+\alpha^2_3)}\\
\alpha_1\alpha_2(\frac{-1}{\alpha^2_3}+\frac{2}{\alpha^2_1+\alpha^2_2+\alpha^2_3})& \frac{\alpha^2_2}{\alpha^2_3} & \frac{-\alpha_2(-\alpha^2_1+\alpha^2_2+\alpha^2_3)}{\alpha_3(\alpha^2_1+\alpha^2_2+\alpha^2_3)} \\
\frac{-\alpha_1(\alpha^2_1-\alpha^2_2+\alpha^2_3)}{\alpha_3(\alpha^2_1+\alpha^2_2+\alpha^2_3)} & \frac{-\alpha_2(-\alpha^2_1+\alpha^2_2+\alpha^2_3)}{\alpha_3(\alpha^2_1+\alpha^2_2+\alpha^2_3)} & 1
\end{array} \right).
\end{eqnarray}
By comparing the 2-3 block of the above structure with the mass
structure in Eq. 32, one can see the constant $a$ is equivalent to
the quantity
($(-\alpha^2_1+\alpha^2_2+\alpha^2_3)/(\alpha^2_1+\alpha^2_2+\alpha^2_3)$)
which is equal to one in the limit $\alpha_1\rightarrow 0$. So, let
us expand the eigenvalues of the right handed neutrino mass
structure in Eq.34 around $\alpha_1$.
\begin{eqnarray}
M_{R1}&=&1+\frac{\alpha^2_2}{\alpha^2_3}+\frac{\alpha^2_1(\alpha^4_2-6\alpha^2_2\alpha^2_3+\alpha^4_3)}{\alpha^2_3(\alpha^2_2+\alpha^2_3)^2}+\mathcal{O}(\alpha_1^4)\nonumber,\\
M_{R2}&=&\frac{4\alpha^2_1\alpha^2_2}{(\alpha^2_2+\alpha^2_3)^2}-\frac{8\alpha^3_1\alpha^3_2\alpha_3}{(\alpha^2_2+\alpha^2_3)^{7/2}}+\mathcal{O}(\alpha_1^4),\\
M_{R3}&=&\frac{4\alpha^2_1\alpha^2_2}{(\alpha^2_2+\alpha^2_3)^2}+\frac{8\alpha^3_1\alpha^3_2\alpha_3}{(\alpha^2_2+\alpha^2_3)^{7/2}}+\mathcal{O}(\alpha_1^4)\nonumber.
\end{eqnarray}
One can see two of the right handed neutrino masses are
approximately degenerate for small values of $\alpha_1$ (i.e.
$M_{R2}\approx M_{R3}$). By setting ($\alpha_1$, $\alpha_2$,
$\alpha_3$, $\Lambda$)=($-0.05$, 0.125, 0.994, $8.42\times10^{15}$),
the numerical fit for the neutrino mixing angles, the light neutrino
masses and right handed neutrino masses are obtained as follows:
\begin{align}
m_1&=0\textrm{ eV},       &\sin\theta_{sol}&=0.551,    &M_{R1}&=8.57\times10^{15}\textrm{ GeV}\nonumber,\\
m_2&=0.01\textrm{ eV},   &\sin\theta_{atm}&=0.776,    &M_{R2}&=1.3\times10^{12}\textrm{ GeV}\nonumber,\\
m_3&=0.056\textrm{ eV},
&\sin\theta_{13}&=0.154,&M_{R3}&=1.28\times10^{12}\textrm{ GeV}.
\end{align}
\section{Conclusion}

Both largeness of neutrino mixing angles and the smallness of CKM
mixing angles can be accounted for by a lopsided structure of the
charged leptons and down quarks. This structure was obtained in
several studies and has been used for many models of fermion masses.
The model that has been studied here is the first SUSY grand
unification model based on the gauge symmetry SO(10) with the
discrete family symmetry $A_4$ that leads to the doubly lopsided
structure. A few works on SO(10) $\times A_4$ have recently been
published, but what makes this work unique is the assumption of
using the minimal set of Higgs fields that break SO(10) to the SM
group. This assumption acts as an important guide for searching for
good models. The possibilities of renormalizable Yukawa interactions
for quarks and leptons are very limited because the minimum Higgs
breaking scheme is imposed and the superpotential must respect the
assigned symmetry of the model. Based on that, a general mass
structure for the heavy generations has been obtained which explains
the following features: (1) $m^0_b\approx m^0_{\tau}$, (2) $\frac
{m^0_{\mu}}{m^0_s}=3$, (3)
$\frac{m^0_c}{m^0_t}<<\frac{m^0_s}{m^0_b}$. It is important to
mention that another work \cite{b} got the same mass structure for
heavy fermions. In that work, the authors did not employ the flavor
symmetry and showed that the hierarchy between the second and third
generations can be understood by choosing the specific direction of
$\langle45_H\rangle $. Also, they employed another adjoint Higgs
field $\langle45_H\rangle $ to include the first family to their
model. On the other hand, in this presented study, the above
features of heavy fermions have been obtained by picking specific
direction of $\langle45_H\rangle $, but the hierarchy between the
three generations can be understood in the framework of
$A_4$-symmetry. Without adding another adjoint to the model, the
first family is successfully included to the model and excellent
predictions are obtained.

For fitting purposes,  some approximate analytical expressions given
in Eq. 26 are derived for mass ratios and mixing angles of the
quarks and the leptons  by combining the Yukawa couplings
represented by three Feynman diagrams in Figures 1-3. However, exact
numerical fitting at low scale was done. Without including the
neutrino sector, the model predictions at low scale for the masses
and the mixing angles (except the atmospheric angle) of the quarks
and the charged leptons, as well as CP violation parameter are in
excellent agreement with data (i.e. within 2$\sigma$). The
atmospheric angle needes to be corrected by considering the neutrino
sector. The symmetry of the model succeeds to produce the
appropriate right handed neutrino structure that gives not only the
correct contribution to the atmospheric angle, but also the correct
neutrino mass differences. The neutrino contribution to the solar
angel is negligible. Besides, after combining the two contributions
(neutrinos and charged leptons contributions) the neutrino mixing
angle $\theta_{13}$ is predicted.
\section*{Acknowledgements}
I would like to thank Prof. K. S. Babu for advice and for reading
the manuscript. This work is supported in part by US Department of
Energy, Grant Numbers DE-FG02-04ER41306 and DE-FG02-ER46140

\newpage

\begin{figure}
\centering
\includegraphics[scale=1]{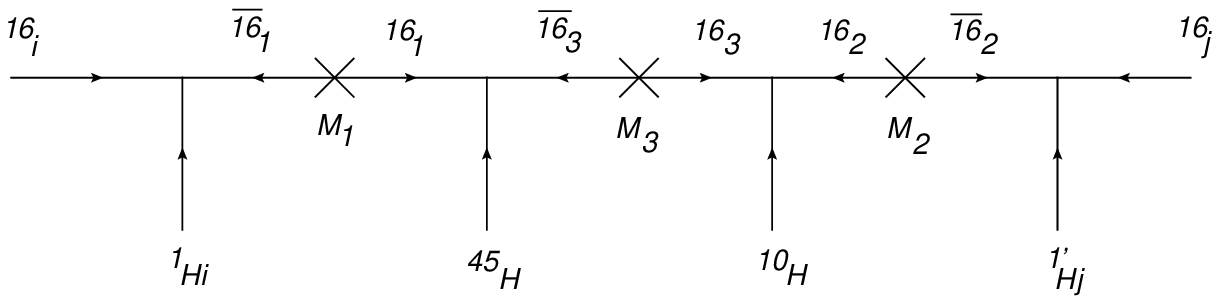}
\caption{ This figure shows diagrammatically the couplings in the
superpotential $W_1$. }
\end{figure}

\begin{figure}
\centering
\includegraphics[scale=1]{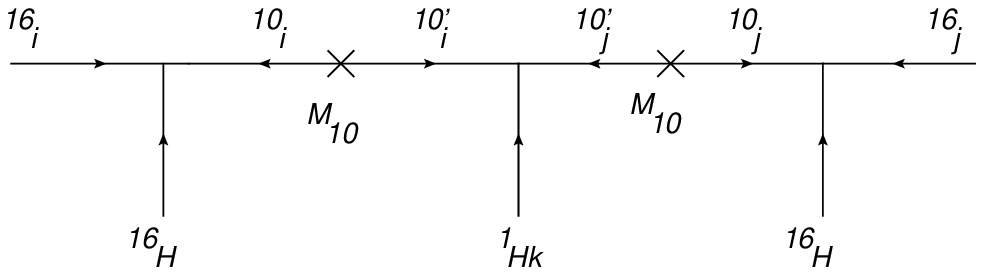}
\caption{This figure leads to the flavor symmetric contribution to
the down quarks and charged leptons. }
\end{figure}

\begin{figure}
\centering
\includegraphics[scale=1]{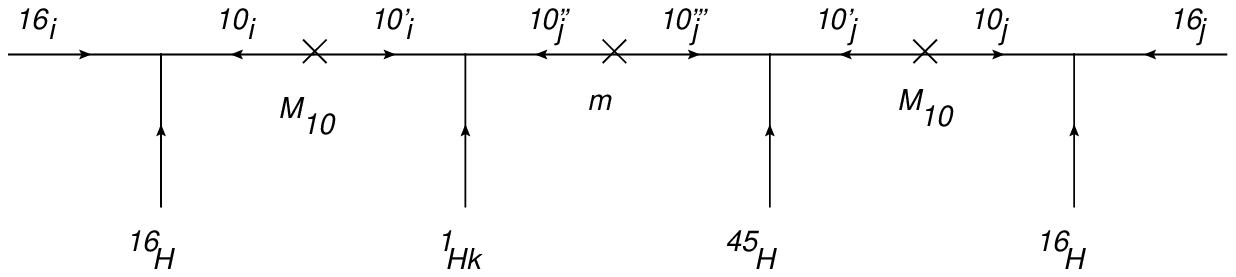}
\caption{This figure leads to the flavor antisymmetric contribution
to the down quarks and charged leptons.}
\end{figure}

\begin{figure}
\centering
\includegraphics[scale=1]{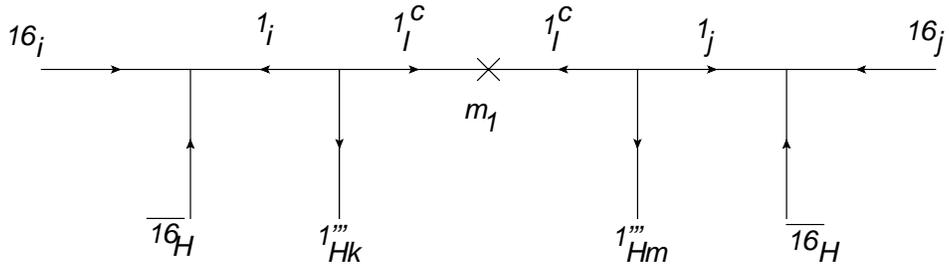}
\caption{This figure leads to the right handed neutrino mass
matrix.}
\end{figure}
\end{document}